\documentclass[a4paper,11pt]{article}

\usepackage{comment}
\usepackage{jinstpub} 
\usepackage{textcomp}
\usepackage{xcolor}

\usepackage{epsfig}
\usepackage{epstopdf}
\usepackage{color} 
\usepackage{url} 

\usepackage{lineno}
\title{\boldmath A first test of CUPID prototypal light detectors with NTD-Ge sensors in a pulse-tube cryostat}

\collaboration{%
\includegraphics[height=17mm]{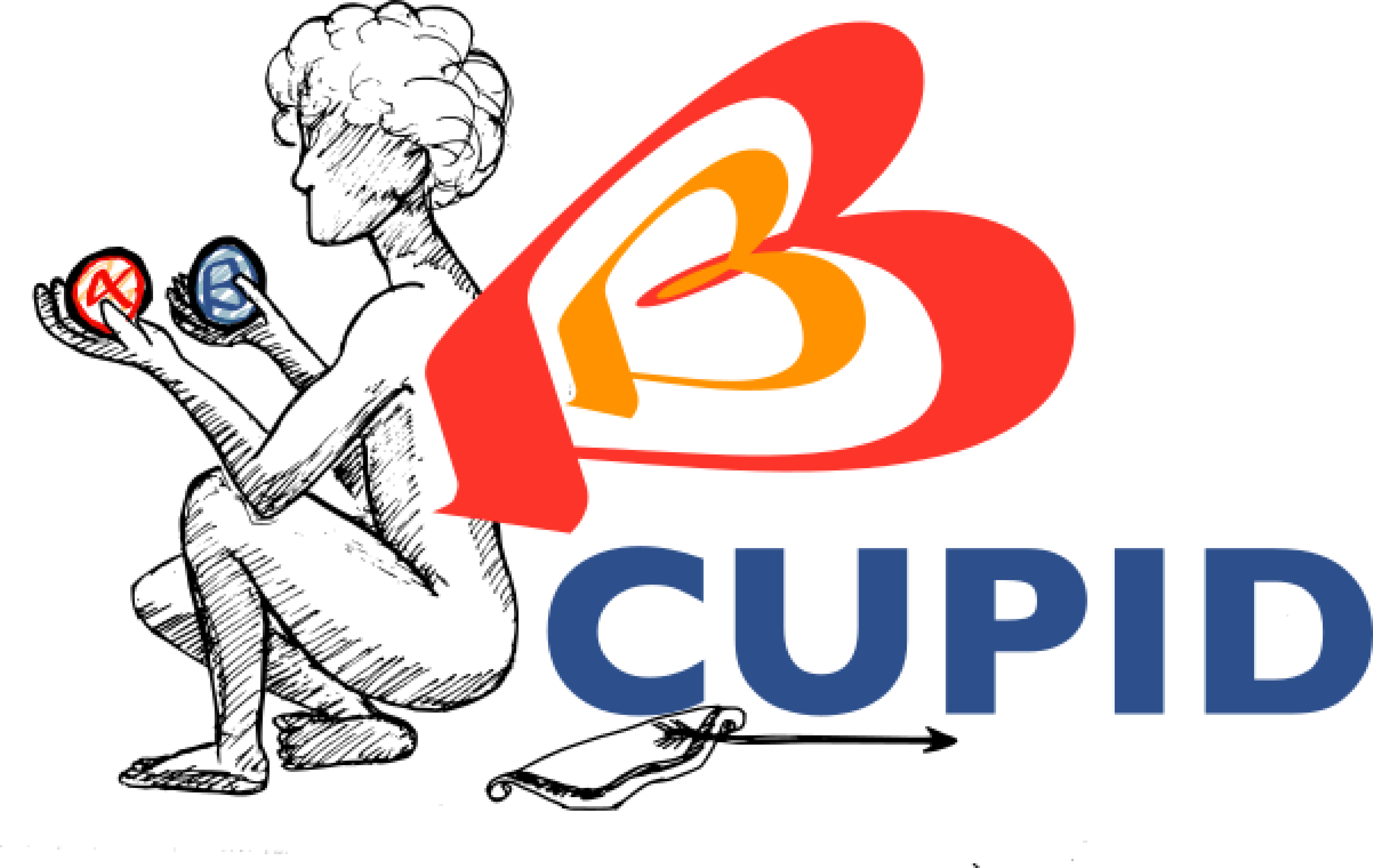}\\[6pt]
CUPID collaboration}

\author[a]{K.~Alfonso,}
\author[b]{A.~Armatol,}
\author[c]{C.~Augier,}
\author[d]{F.T.~Avignone~III,}
\author[e]{O.~Azzolini,}
\author[f]{M.~Balata,}
\author[g]{A.S.~Barabash,}
\author[h]{G.~Bari,}
\author[i,j]{A.~Barresi,}
\author[b]{D.~Baudin,}
% ...

\affiliation[a]{Virginia Polytechnic Institute and State University, Blacksburg, VA, USA}
\affiliation[b]{IRFU, CEA, Université Paris-Saclay, Saclay, France}
\affiliation[c]{Univ Lyon, Université Lyon 1, CNRS/IN2P3, IP2I-Lyon, Villeurbanne, France}
\affiliation[d]{University of South Carolina, Columbia, SC, USA}
\affiliation[e]{INFN Laboratori Nazionali di Legnaro, Legnaro, Italy}
\affiliation[f]{INFN Laboratori Nazionali del Gran Sasso, Assergi, AQ, Italy}
\affiliation[g]{National Research Center Kurchatov Institute, Kurchatov Complex of Theoretical and Experimental Physics, Moscow, Russia}
\affiliation[h]{INFN Sezione di Bologna, Bologna, Italy}
\affiliation[i]{INFN Sezione di Milano-Bicocca, Milan, Italy}
\affiliation[j]{University of Milano-Bicocca, Milan, Italy}
% ...

\emailAdd{cupid.publications@lngs.infn.it}

\abstract{
CUPID is a next-generation bolometric experiment aiming at searching for neutrinoless double-beta decay with ~250 kg of isotopic mass of $^{100}$Mo. It will operate at $\sim$10~mK in a cryostat currently hosting a similar-scale bolometric array for the CUORE experiment at the Gran Sasso National Laboratory (Italy). CUPID will be based on large-volume scintillating bolometers consisting of $^{100}$Mo-enriched Li$_2$MoO$_4$ crystals, facing thin Ge-wafer-based bolometric light detectors. In the CUPID design, the detector structure is novel and needs to be validated. In particular, the CUORE cryostat presents a high level of mechanical vibrations due to the use of pulse tubes and the effect of vibrations on the detector performance must be investigated. In this paper we report the first test of the CUPID-design bolometric light detectors with NTD-Ge sensors in a dilution refrigerator equipped with a pulse tube in an above-ground lab. Light detectors are characterized in terms of sensitivity, energy resolution, pulse time constants, and noise power spectrum. Despite the challenging noisy environment due to pulse-tube-induced vibrations, we demonstrate that all the four tested light detectors comply with the CUPID goal in terms of intrinsic energy resolution of 100 eV RMS baseline noise. Indeed, we have measured 70--90 eV RMS for the four devices, which show an excellent reproducibility. We have also obtained outstanding energy resolutions at the 356 keV line from a $^{133}$Ba source with one light detector achieving 0.71(5)~keV FWHM, which is ---to our knowledge--- the best ever obtained when compared to $\gamma$ detectors of any technology in this energy range. 
}


\begin{document}

\maketitle
\flushbottom

%%============================================================================================
\section{Introduction}
\label{sec:intro}

Neutrinoless double-beta decay ($0\nu2\beta$) is a hypothetical nuclear process energetically allowed for a tens of even-even nuclei \cite{Tretyak:2002}, that can occur beside the rare but Standard-Model-allowed two-neutrino double-beta decay ($2\nu2\beta$) \cite{GoeppertMayer:1935}. While in the latter two electrons and two anti-neutrinos are produced, the former implies the emission of two electrons only. The $2\nu2\beta$ process has been observed in several isotopes with the half-lives ranging between 10$^{19}$--10$^{24}$~yr \cite{Pritychenko:2023,Barabash:2020,Belli:2020a,Saakyan:2013}. On the other hand, a possible existence of $0\nu2\beta$, proposed by W. H. Furry in 1939 \cite{Furry:1939} and never detected, is still under probe by many experiments; the most stringent half-life limits are at the level of 10$^{24}$--10$^{26}$~yr   \cite{Arnold:2015,Anton:2019,Adams:2020,Agostini:2020,Azzolini:2022,Augier:2022,Arnquist:2023,Abe:2023}. Its observation would establish that neutrinos are Majorana particles (i.e., they are their own antiparticles) \cite{Schechter:1982}. In addition, its existence would require an extension of the Standard Model to include lepton number violation \cite{Agostini:2022,Adams:2022b,Dolinski:2019,DellOro:2016tmg,Vergados:2016,Pas:2015,Vergados:2012,Giuliani:2012a,Rodejohann:2012,Deppisch:2012,Rodejohann:2011}. 

Experiments searching for $0\nu2\beta$ are looking for a peak at the $Q$-value ($Q_{\beta\beta}$) of the transition, corresponding to the sum of the kinetic energies carried by the two emitted electrons. There are 35 nuclei that can energetically undergo double-beta decay \cite{Tretyak:2002}, however not all of them are viable for experimental probing \cite{Giuliani:2012a,GomezCadenas:2012,GomezCadenas:2014,Cremonesi:2014,DellOro:2016tmg,Poda:2017}. If an appropriate detector technology is available, the most favourable isotopes to study are those with a $Q_{\beta\beta}$ above the natural $\gamma$ radioactivity, that breaks off at the 2615 keV line of $^{208}$Tl, with the exception of rare emissions with negligible branching ratio. This implies less background in the region of interest (ROI) and a higher decay probability, since the phase-space of the process increases sharply with $Q_{\beta\beta}$. 

The decay can be investigated with low-temperature calorimeters, designated also as bolometers, which embed the double-beta decay isotope under study and detect particle interactions via a small temperature rise read out by a dedicated thermometer \cite{Fiorini:1983,Pirro:2017,Poda:2017,Bellini:2018,Biassoni:2020,Poda:2021,Zolotarova:2021}. Bolometers are among the most promising methods to search for $0\nu2\beta$ thanks to their excellent energy resolution and high detection efficiency. They can be made from a wide choice of absorber materials having a high internal radiopurity. In addition, bolometers can be scaled up to large detector mass through arrays, as demonstrated by CUORE, which is the first tonne-scale bolometric experiment and has been taking data at the Gran Sasso National Laboratory in Italy since 2017 \cite{Alduino:2018,Adams:2022a,Adams:2022}. CUORE searches for the $0\nu2\beta$ decay of $^{130}$Te ($Q_{\beta\beta}$ = 2528 keV) that is contained in $^{\mathrm{nat}}$TeO$_2$ crystals. 

The sensitivity of CUORE is limited by the $\alpha$ background ---responsible for a background index of $\sim$10$^{-2}$ counts/(keV$\cdot$kg$\cdot$yr) in the ROI--- from radioactive surface contamination of the surrounding materials \cite{Alduino:2017pni,Alduino:2017}. CUPID (CUORE Upgrade with Particle IDentification) is the proposed successor of CUORE with a technique to reduce the background contribution from $\alpha$ surface contamination \cite{Armstrong:2019inu,Wang:2015raa}. It will be based on scintillating bolometers using Li$_2$MoO$_4$ crystals \cite{Barinova:2010,Cardani:2013,Bekker:2016} as a sensitive material, embedding the double-beta decay isotope $^{100}$Mo and coupled to cryogenic Ge light detectors (LDs) \cite{Armstrong:2019inu} that work as bolometers as well. Bolometric LDs are known for good energy resolution and low threshold, and are easier to operate than conventional ones in a millikelvin environment. Scintillating bolometers exploit the fact that $\alpha$ and $\beta/\gamma$ have different scintillation light yields, making an efficient rejection of the dominant $\alpha$ background \cite{Pirro:2006,Poda:2021}. The isotope $^{100}$Mo fulfills the major requirement of having a high $Q_{\beta \beta}$ above 2615 keV (at 3034~keV). A technology of $^{100}$Mo-enriched Li$_2$MoO$_4$ scintillating bolometers was developed in the LUMINEU project, showing a high energy resolution in the ROI ($\sim$5 keV FWHM), high internal radiopurity, and sufficient light yield to reject $\alpha$ particles \cite{Armengaud:2017,Grigorieva:2017,Poda:2017a}. The two demonstrators of CUPID ---CUPID-0 \cite{Azzolini:2018tum,Azzolini:2019tta,Azzolini:2022} and CUPID-Mo \cite{Armengaud:2020a,Armengaud:2021,Augier:2022}, based on the scintillators Zn$^{82}$Se and 
Li$_2$$^{100}$MoO$_4$ respectively--- have successfully validated the dual heat-light readout technology that will be adopted in CUPID. In contrast to the demonstrators, which use cylindrical crystals and circular-shape LDs, CUPID will operate cubic Li$_2$MoO$_4$ crystals \cite{Armatol:2021b,Armatol:2021a,Alfonso:2022,CrossCupidTower:2023a}, requiring the LD shape to be adapted to the crystal face to increase scintillation photons collection efficiency~\cite{Armatol:2021a,CrossCupidTower:2023a}.

The CUPID detector configuration proposes a novel mechanical structure to hold the crystals and couple them to the LDs \cite{Alfonso:2022}. In this work we will describe how the LDs were assembled and present results on their performance in a dilution refrigerator with pulse-tube cooling (as foreseen in CUPID).

%%============================================================================================
\section{Assembly and bolometric test}

\begin{figure}[t]
\centering % \begin{center}/\end{center} takes some additional vertical space
\includegraphics[width=0.5\textwidth]{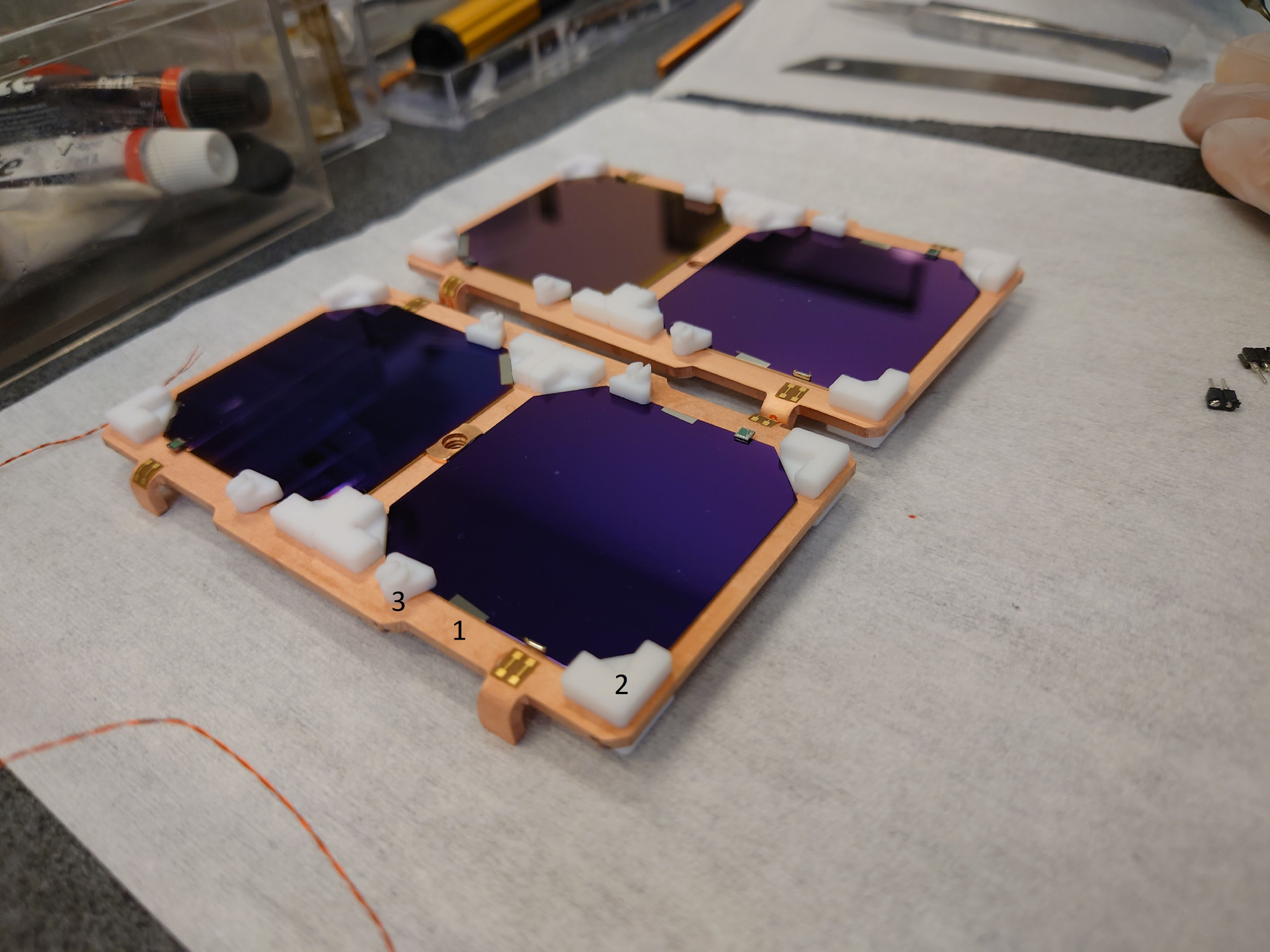}
\includegraphics[width=0.4\textwidth]{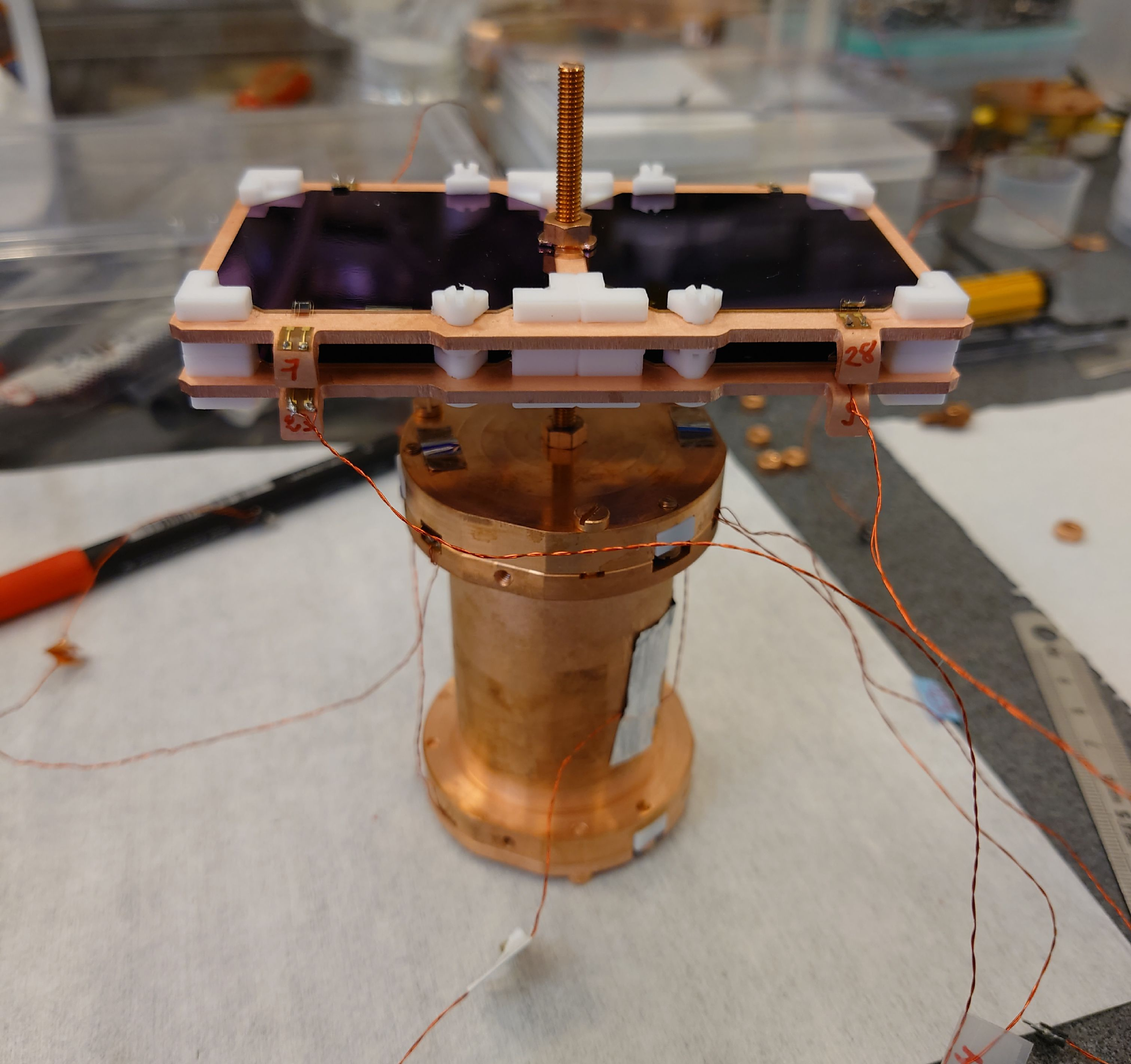}
%\qquad
%\includegraphics[width=.40\textwidth,trim=200 200 200 0,clip]{DSC_1677.pdf}
%\includegraphics[width=.25\textwidth,trim=200 200 200 0,clip]{DSC_1677.pdf}
\caption{Left: Four Ge LDs assembled following the CUPID baseline tower design (a rendering of a single module can be found in \cite{Alfonso:2022}). Each LD is resting directly on the Cu holder (1) and surrounded by four PTFE pieces (2). Two additional PTFE pieces (3) are also used to gently clamp them. An NTD Ge thermistor and a heater are glued on each LD. Right: An array of four LDs ready to be installed in a pulse-tube cryostat.} 
\label{fig:LDs}
\end{figure}

The assembly procedure of CUPID \cite{Alfonso:2022} has been simplified compared to CUORE \cite{Alduino:2016b,Buccheri:2014} and to the demonstrators CUPID-0 \cite{Azzolini:2018tum} and CUPID-Mo \cite{Armengaud:2020a}. A CUPID tower will consist of 14 stacked modules. Each module will comprise a copper frame hosting two LDs and two cubic Li$_2$MoO$_4$ crystals 45 mm side each. The innovation consists in the fact that crystals are not firmly clamped by polytetrafluoroethylene (PTFE) elements but, during the assembly, they just rest by gravity on four PTFE pieces, which are in turn secured on the copper frame and surrounding the LD Ge wafers. The copper frame and the LDs constitute an independent element, that is shown in figure~\ref{fig:LDs}, left panel. The tower is formed by just piling alternatively the copper frames (with their LDs) and the crystals. At the end, a vertical copper skeleton frame is fixed between lowest and top frames to keep the floors together. The crystals do not exert any pressure on the LD Ge wafers, as the latter are protected by the four PTFE elements. A pressure on the top of the tower can be applied by a copper screw (a part of the skeleton frame) on a copper plate placed at the top in order to stabilize the tower. This simplified structure has some advantages: 
a) less time and effort to assemble; 
b) simplification in the production of copper materials; 
c) absence of screws and threads, which implies easier surface cleaning and no risk for stuck screws during the assembly; 
d) less passive materials; 
e) self-adjustment of the tower under thermal contractions during the cooling down. 
The design is also characterized by having an open structure, i.e., the crystals' surfaces will be facing each other with no material in between, as implemented in CUORE \cite{Adams:2021}. This is particularly important to perform coincidence analysis for background mitigation, so that events in which particles share their energy between two nearby crystals can be rejected.

Figure 1 (left) shows the assembled LDs used for this work. They are based on octagonal-shape 500~$\mu$m-thick high-purity Ge wafers (resistivity higher than 50~$\Omega \cdot$cm at room temperature), resting on a copper holder and kept between four PTFE pieces and locked using two PTFE rotating pieces. The wafers were coated with an anti-reflecting SiO layer \cite{Mancuso:2014} with a thickness of 60 nm using two methods: evaporation (performed in IJCLab, Orsay, France) and sputtering (performed in Argonne National Laboratory, US). Two of each kind were chosen for the test. The purpose of the SiO coating is to increase the light absorption by the LDs \cite{Mancuso:2014}; it has been already validated in CUPID-0 \cite{Azzolini:2018tum} and CUPID-Mo \cite{Armengaud:2020a}. Each wafer carries a neutron-transmutation-doped Ge thermistor \cite{Haller:1994} (designated simply as NTD in the following) with a size of 3$\times$0.5$\times$1~mm$^3$ and a heater consisting of a P-doped Si chip  \cite{Andreotti:2012}. The NTD collects the phonons produced in the absorber after energy deposition and converts them into an electrical signal. The heater is typically exploited to correct for the thermal gain drift of the bolometer by delivering a constant amount of energy by Joule effect periodically \cite{Alessandrello:1998}, but it was not used in the present work. Both chips are glued on the wafer by means of an epoxy glue (Araldite\textregistered~Rapid). The tested modules had Kapton\textregistered~foil with gold contacts used to wire-bond the NTD and heater for the electrical readout. 
% Azzolini:2019,Azzolini:2019tta

This design provides a different thermal and mechanical configuration of securing the Ge wafers compared to the ways adopted in CUPID-0 and CUPID-Mo. In fact, the wafers are in direct contact with the copper frame, that is at base temperature, and are less firmly fixed. This could affect the bolometric performance, due to a stronger thermal coupling to the bath with respect to the previous experiments, and induce some vibrational noise. 

This work is a test of the bolometric performance of the LDs in the new holder. There was no crystal scintillator facing the Ge wafer to evaluate the light yield registered in the LDs. The focus will be on characterizing these devices by looking at the baseline resolution and the time characteristics of the thermal pulses, which is essential for pile-up rejection capability \cite{Chernyak:2012,Chernyak:2014,Chernyak:2017,CROSSpileup:2023}. In fact, different holder designs can lead to different noise power spectrum, and the time characteristics (rise-time and the decay-time) are connected to the thermal coupling of the LD with the holder. These LDs ---when tested together with Li$_2$MoO$_4$ scintillating crystals in the Gran Sasso laboratory--- showed a light collection efficiency of around 0.33 keV/MeV for $\gamma$($\beta$) radiation \cite{Alfonso:2022}. This means that when $\beta$ or $\gamma$ releases 1~MeV in the scintillating crystal the corresponding light collected by the bolometric photodetector induces a signal of $\sim$0.33~keV. Thus, a $0\nu2\beta$ decay signal ($\sim$3~MeV) corresponds to a $\sim$1~keV pulse in the LD. This allows us to estimate the signal-to-noise ratio once we know the baseline width in energy.

The assembled LDs were tested in a pulse-tube dry cryostat at IJCLab (France), housing a dilution unit capable of reaching a base temperature around 12 mK \cite{Mancuso:2014a}. The two copper frames carrying the LDs were placed on top of each other (figure~\ref{fig:LDs}, right) and screwed to a spring-suspended plate in the cryostat experimental volume. The floating plate is thermally connected to the coldest point of the dilution unit ---the mixing chamber--- by a soft copper link and is suspended by four stainless-steel springs to reduce the vibrations coming from the cryostat body. This suspension system is characterized by a vertical oscillation frequency of $\sim$3~Hz and a horizontal (pendulum) oscillation frequency of $\sim$1~Hz. A pulse-tube cryocooler is used to cool down the 80~K and 4~K thermal stages of the cryostat avoiding the use of cryogenic liquids. This part of the refrigeration system is a dominant source of vibrational noise for bolometers, particularly for LDs \cite{Olivieri:2017,Armengaud:2017}. The CUORE cryostat \cite{Alduino:2019}, which will host CUPID, is equipped with five operational pulse tubes (four of them are used during the CUORE data taking), but no LD has ever been tested there. We consider therefore the present experiment as a preliminary but relevant validation of the CUPID LDs and their mechanical structure in a challenging environment in terms of vibrations.

%%============================================================================================
\section{Data taking and processing}

In an NTD-based bolometer, as implemented in CUORE and its predecessors (CUORE-0, Cuoricino, Mi-DBD \cite{Brofferio:2018}), CUPID-0, and CUPID-Mo, the phonon signal is converted to an electrical pulse by flowing a constant current in the high-resistance (M$\Omega$ scale) thermistor and registering the voltage changes across it. The current is injected through a load resistor, whose resistance (G$\Omega$ scale) is much higher than the NTD resistance at the operation point. Low noise, DC-coupled, voltage-sensitive amplifiers are used to read out the pulses. In the present experiment, the voltage output of the thermistor ---acquired using a predecessor of the CUORE front-end electronics \cite{Arnaboldi:2002}--- is streamed and stored continuously. We used a DAQ software to save on disk the data stream acquired through a 16 bit analog-to-digital board with a sampling rate of 5 kS/s. The data are then processed offline using a MATLAB-based analysis tool that was developed at IJCLab \cite{Mancuso:2016}. The program extracts the signal amplitudes applying the Gatti-Manfredi optimum filter \cite{Gatti:1986}, whose purpose is to maximise the signal-to-noise (S/N) ratio in the signal bandwidth, providing the best estimation of the signal amplitude under given noise conditions and with known signal shape. 

The data processing starts by building a template for a measurement, which consists of a mean pulse and of a noise spectrum. The mean pulse (with the signal's amplitude maximum normalized to one) is obtained by averaging over a certain number of pulses, tens--hundreds, chosen by visual inspection within a specific amplitude range (muon bump region in this case). The noise power spectrum is built by exploiting a large number of baseline samples ($\sim$10000) not populated by particle pulses. Then, we applied optimum filtering to trigger events in the data stream. The optimum triggering depends on two parameters: (i) the threshold, an amplitude value above which events are tagged ---it is chosen to be a multiple of the standard deviation of the filtered baseline (typically, 5$\sigma$) to avoid the triggering of noise fluctuations; (ii) a pulse-shape parameter called ``correlation'' (it lies in the range $[0,1]$), which quantifies how much the filtered pulse is similar to the mean pulse. In fact, this parameter is the Pearson's linear correlation coefficient between the individual pulse above threshold and the mean pulse, both at the filter output (if they are identical, the correlation is equal to 1, otherwise it is below 1). To be triggered, a pulse must have a correlation above a given fixed value. 
For each triggered event, the data processing tool computes the amplitude estimated by the optimum filter and a set of pulse-shape parameters; the parameters relevant for the present work are: the correlation, the rise-time (pulse duration from 10$\%$ to 90$\%$ of the maximum amplitude in the rising part), and the decay-time (pulse duration from 90$\%$ to 30$\%$ of the maximum amplitude in the decaying part).

%%============================================================================================
\section{Analysis of light detectors performance}

\begin{figure}
\centering % \begin{center}/\end{center} takes some additional vertical space
\includegraphics[width=0.7\textwidth]{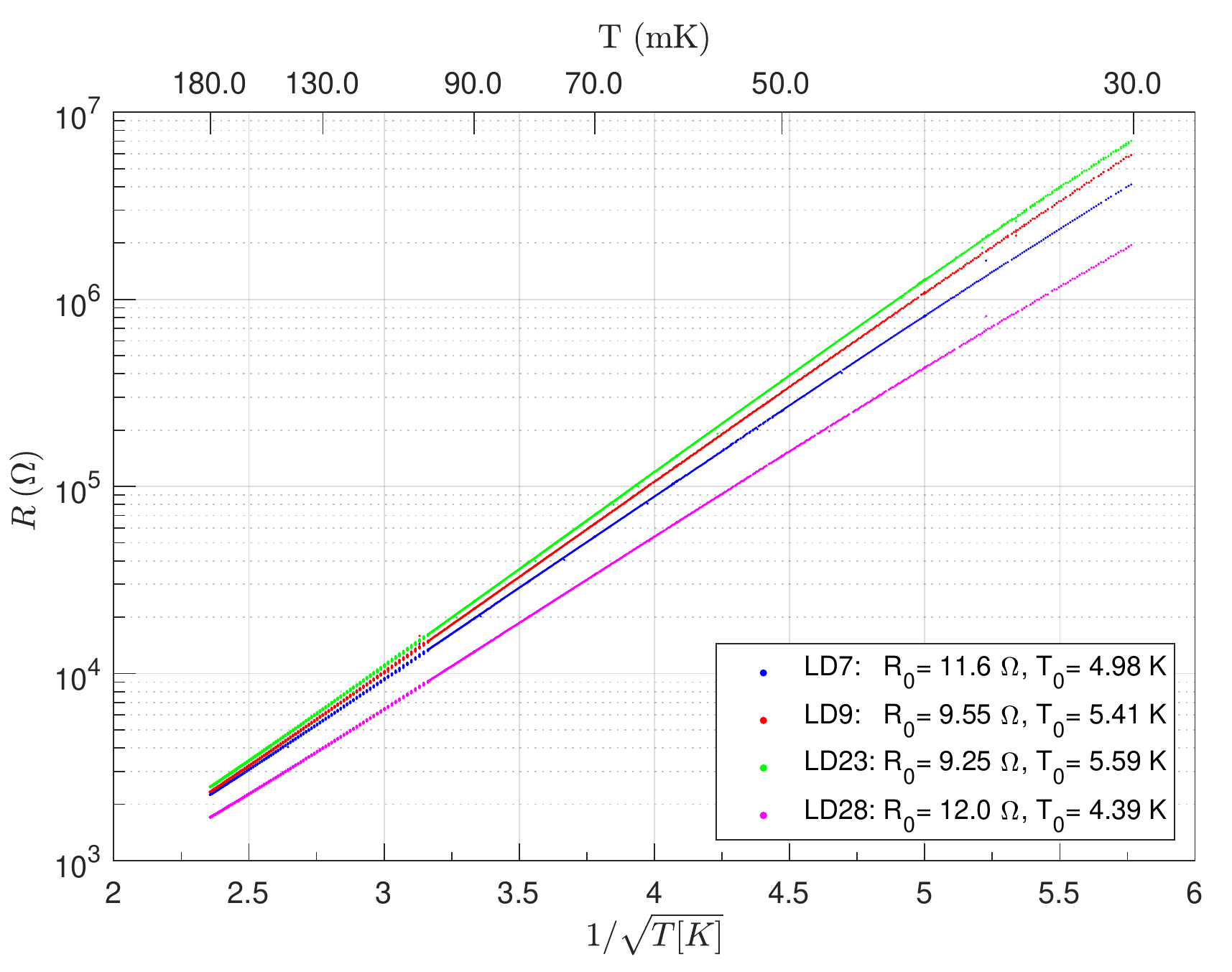}
\caption{Low-temperature $R(T)$ characterization of four NTDs, used to instrument the light detectors in the present study, in terms of $R_0$ and $T_0$ parameters (see text). The values are obtained from the $R(T)$ data in the fit range of 0.03--0.18 K. The characterization was performed with the NTDs already glued at the Ge wafers.}
\label{fig:RT}
\end{figure}  

After reaching the base temperature of the cryostat ($\sim$12 mK), we performed a static characterization of NTDs to map their resistance $R$ as a function of the temperature $T$ using a resistance bridge. In this type of thermistors, the dependence of the resistance on the temperature, corresponding to the variable range hopping conduction regime \cite{Hill:1976}, is given by:
\begin{equation}
    R(T)=R_0 \cdot \exp \left( \sqrt { \frac{T_0}{T} } \right)  , 
    \label{hopping}
\end{equation}

\noindent where $R_0$ (in $\Omega$) depends mainly on the geometry of the thermistor, while $T_0$ (in K) characterizes the impact of the Ge doping and the compensation level and is independent of the thermistor size and the thermal contact geometry.

The $R$($T$) data were acquired in the temperature range of 30--180 mK (below 30~mK the NTDs resistances are too large to be acquired by our resistance bridge). It is convenient to represent such data in log($R$) versus 1/$\sqrt{T}$, to make a linear fit and extract the $R_0$ and $T_0$ parameters. The resulting $R_0$ and $T_0$ values are shown in figure~\ref{fig:RT} (where LDs are indicated with the numbering adopted in the set of 30 LDs prepared for the full tower). We found that the parameter $T_0$ is substantially higher than the one of a typical NTD of the same production batch with a larger size (e.g., 3$\times$1$\times$1~mm$^3$ and 3$\times$3$\times$1~mm$^3$ used in CUPID-Mo \cite{Armengaud:2020a}). This leads to higher resistances at equal temperatures, even when correcting for the corresponding geometrical factor. This could suggest that the $R(T)$ behavior is affected by mechanical stresses induced by the glue on these small elements. It is known that this mechanism tends to increase $T_0$ and consequently the resistance at a given temperature. For example, the typical value for $T_0$ is $\sim 3.8$~K for a non-stressed thermistor with the same doping level as the ones used for these LDs, to be compared with the range 4.4--5.6~K observed here (figure \ref{fig:RT}). 

\begin{figure}
\centering % \begin{center}/\end{center} takes some additional vertical space
\includegraphics[width=0.7\textwidth]{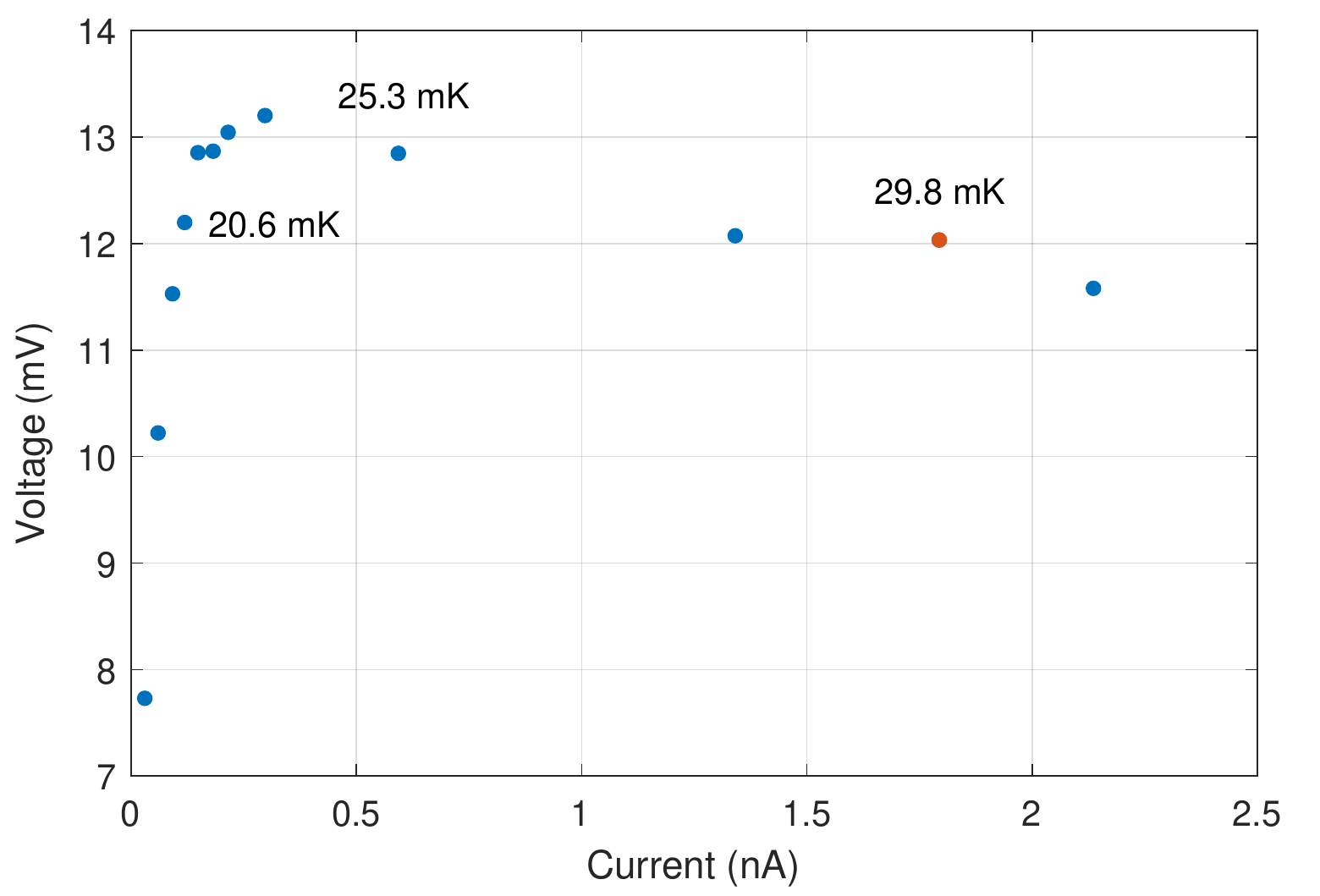}
\caption{The IV curve of LD9  at a floating plate temperature of 15 mK. The plot shows also the temperature of the NTD at difference biases. The red dot represents the optimum working point (the best S/N ratio) that corresponds to an NTD resistance of 6.7 M$\Omega$ (29.8 mK temperature).}
\label{fig:IVcurve}
\end{figure}

\begin{figure}
\centering % \begin{center}/\end{center} takes some additional vertical space
\includegraphics[width=0.7\textwidth]{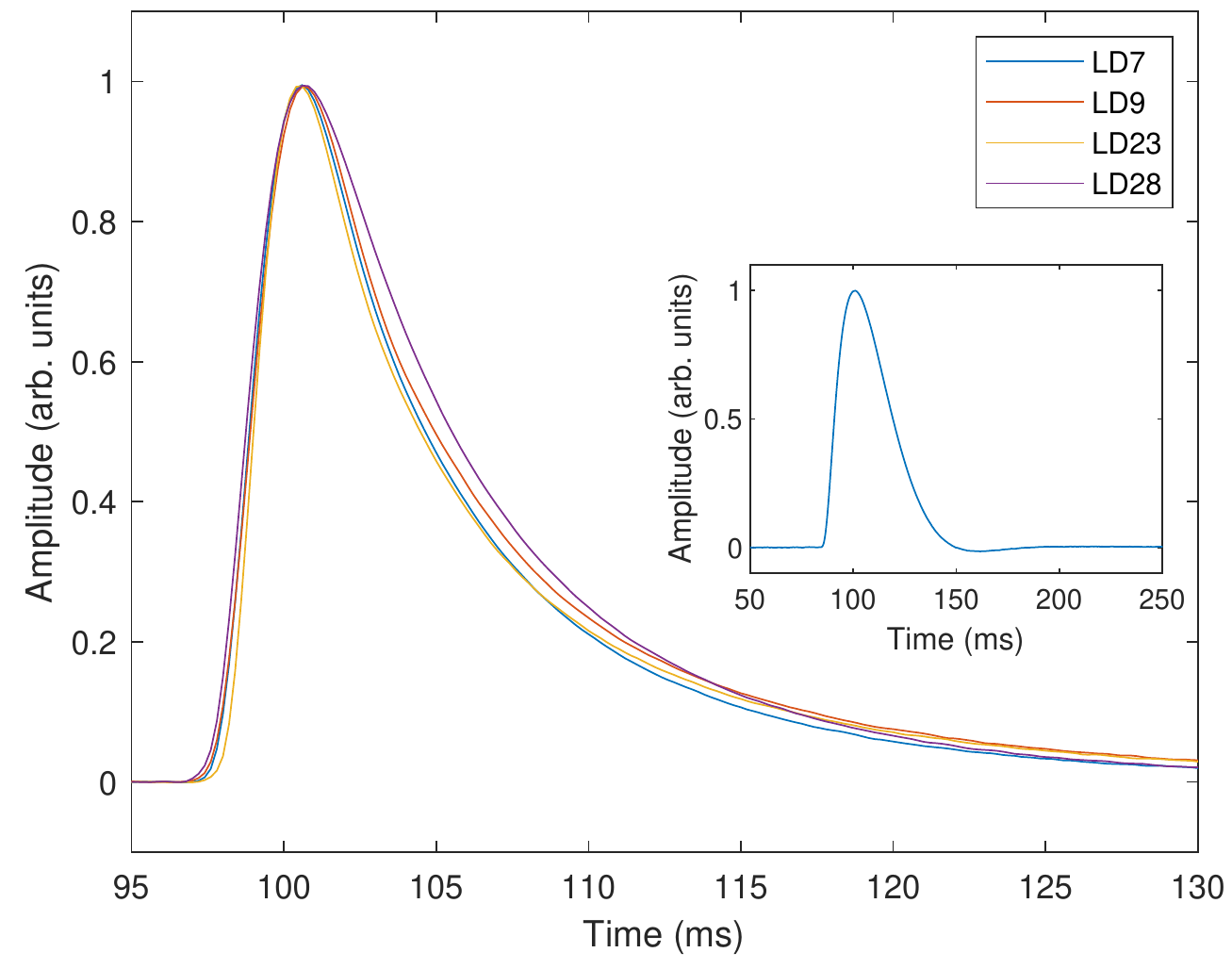}
\caption{Normalized mean pulses of four LDs measured at the optimum working point (corresponding to a high NTD bias, at a few nA current) with a detector plate temperature of 15 mK. \textit{Inset:} Mean pulse of a  LD at low NTD bias (0.1 nA corresponding to 80 M$\Omega$ of the NTD resistance) which shows a comparatively slow signal with a bouncing pulse shape.}
\label{fig:signals}
\end{figure}

For the bolometric test, the detector plate temperature was stabilized at 15~mK. The optimum working points were obtained prior to data taking by choosing the bias current that provides the highest S/N ratio. In order to monitor the signal amplitude during the working point optimization, we used a room-temperature LED (820 nm) to inject periodically a burst of photons through an optic fibre to the experimental cavity, shining the LDs. 
At the optimum working point, where the applied bias currents are in the range 1.5--2 nA, the resistances of the NTDs lay in the interval 3--7~M$\Omega$, corresponding to temperatures around 25--30~mK. Figure \ref{fig:IVcurve} shows an example of a current-voltage (IV) curve of one of the LDs illustrating the best S/N ratio working point. 
As foreseen from the results of the NTD characterization (see figure~\ref{fig:RT}), the NTDs showed very high resistances ($\sim$0.1--1~G$\Omega$) when applying low bias (fraction of nA), accompanied by a noisy baseline and bouncing pulse shapes (illustrated in the inset of figure~\ref{fig:signals}). This oscillating behaviour is often observed in very high-impedance bolometers and is related to the combination of the intrinsically inductive response of the bolometer \cite{Jones:1953,Mather:1982, Alfonso:2021}, its high resistance, and the parasitic electrical capacitance of the readout cables.

Bolometric signals of LDs operated at the chosen working points are shown in figure~\ref{fig:signals}. The rise-times are in the 1.7--2~ms range, about a factor 2 longer than the CUPID goal \cite{Armstrong:2019inu}. Detector speed is important, as a fast LD can be crucial to mitigate the background induced by random coincidences of events from $2\nu2\beta$ decay by rejecting pile-up via pulse-shape discrimination \cite{Chernyak:2012,Chernyak:2014,Chernyak:2017,CROSSpileup:2023}. LDs can be made faster ---keeping the current design--- by acting on the NTD doping level, so that they have lower resistances at the same temperature, and by improving the coupling with the wafer with bonding techniques more efficient than gluing, for example the eutectic process~\cite{Nutini:2021}. Also, a faster response can be achieved by a stronger NTD polarization, with the cost of detector sensitivity and noise degradation \cite{Beeman:2013b}.

\begin{figure}[t]
\centering % \begin{center}/\end{center} takes some additional vertical space
\includegraphics[width=0.8\textwidth]{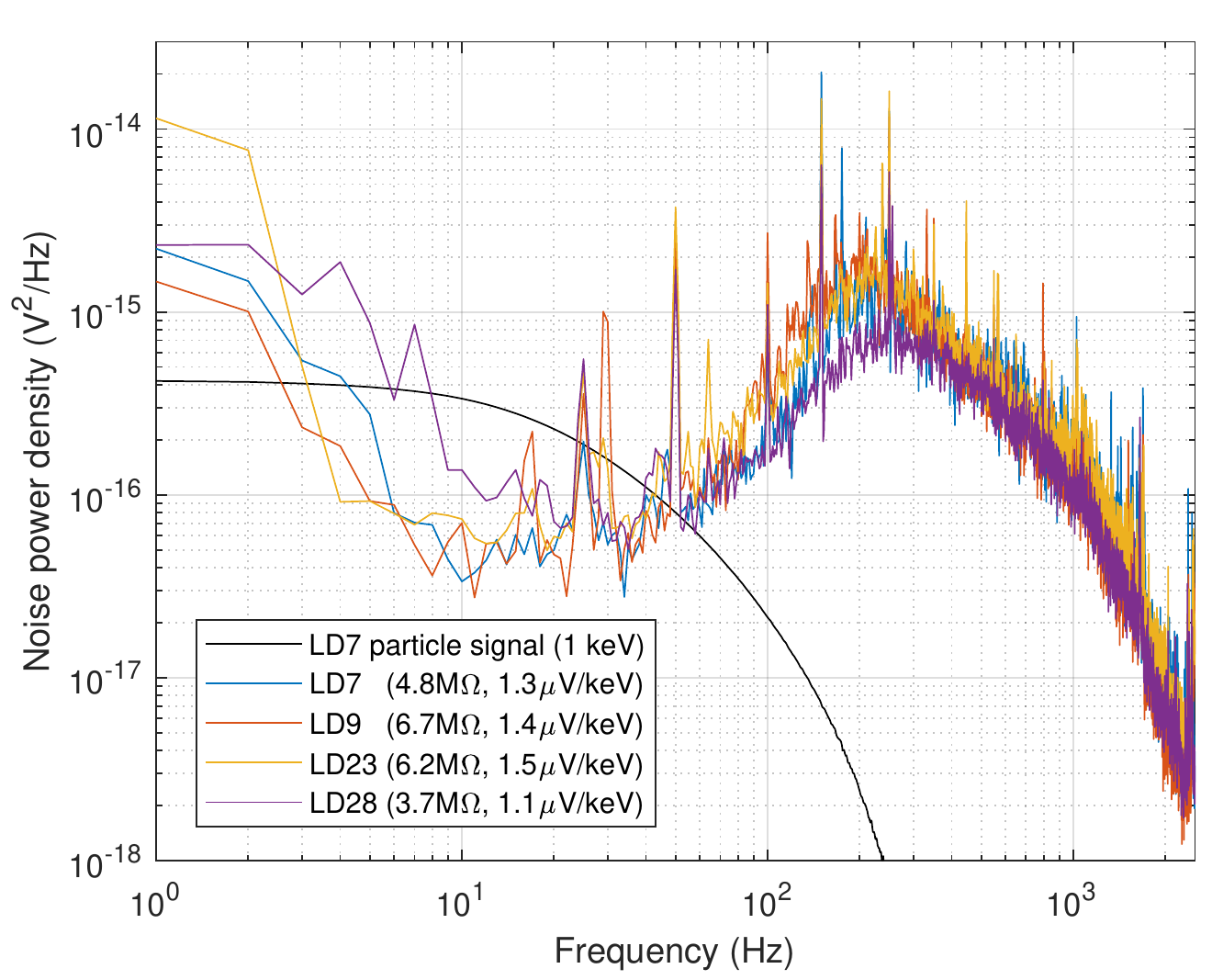}
\caption{The noise power density spectra of the LD readouts compared with a signal of 1~keV in the frequency domain for one detector. This amplitude corresponds to a light signal associated to a full $Q_{\beta\beta}$ energy release. To obtain the frequency representation of the signal, we take the square of the module of its Fourier transform, which has units V$^2$/Hz$^2$. Then, we multiply it by the frequency bin in order to obtain the same units as the noise power density (V$^2$/Hz). The useful bandwidth is approximately 5--50 Hz. The low frequency noise spoils significantly the LD performance. Some peaks associated to vibrations are visible in the 10--50 Hz range. The Bessel cutoff frequency is 675~Hz. The computation of the sensitivities (in $\mu$V/keV) given in the legend is explained in the text.}
\label{fig:noise}
\end{figure}

The noise power density spectra at the working point for each LD are shown in figure~\ref{fig:noise}, compared with a typical signal in the frequency domain. Below 10 Hz, we observe that the noise power spectra increase as the frequency decreases. We have noticed that this effect is smaller when we switch temporarily the pulse tube off, showing that this rise is related to vibrations. The 50~Hz peak and its harmonics are related to the electrical configuration external to the cryostat. Other peaks appear in the region 10--50~Hz; they are connected to vibration modes of the cryostat and cabling and do not change when the pulse tube is switched off. The effect of the noise peaks can be mitigated during data processing using the optimum filter. We remark that the total intrinsic noise (amplifier input noise as well the Johnson noise from the NTD thermistor and the load resistor) is expected to be in the range (3--4)$\times10^{-17}$~V$^2$/Hz. Only in the region 10--40~Hz, the noise spectra approach these values (see figure~\ref{fig:noise}), showing that there is still a wide margin of improvement by reducing the spurious noise sources, specially related to vibrations.

\begin{figure}[t]
\centering % \begin{center}/\end{center} takes some additional vertical space
\includegraphics[width=0.7\textwidth]{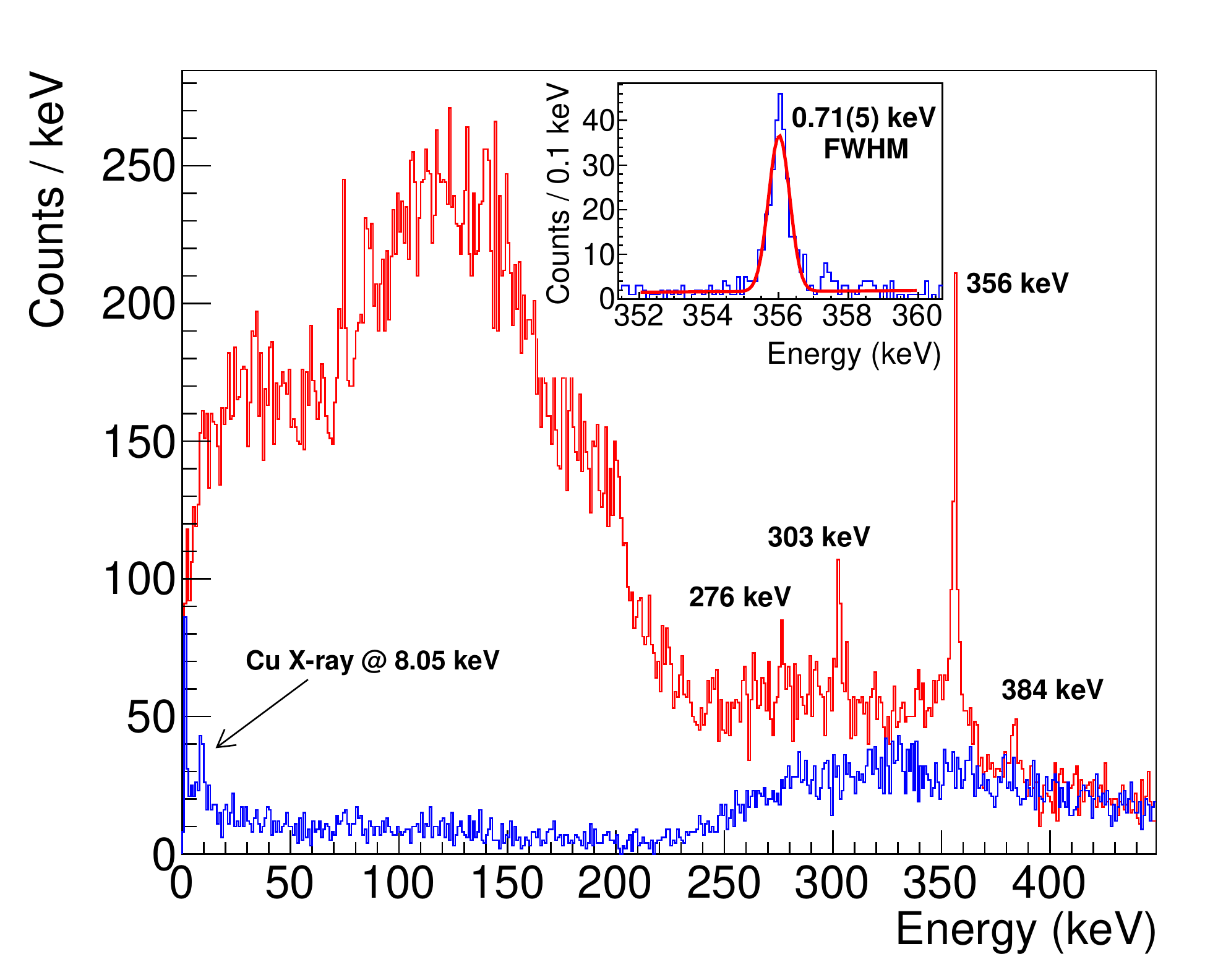}
\caption{Comparison of the LD9 energy spectra acquired with (red; 9.3 h) and without (blue; 9.3 h) a $^{133}$Ba source, which was placed outside the cryostat. The source data exhibit several peaks, which correspond to the most intense $\gamma$ quanta emitted by the $^{133}$Ba source. A bump of events, seen in both data at $\sim$200--500 keV and peaked at $\sim$320~keV, is originated from cosmic muons passing through the Ge wafer. Particle interactions with copper parts of the detector holder and the screen around it can induce Cu X-ray fluorescence detectable by the LD. \textit{Inset:} A fit with a simple model (Gaussian together with a linear background component) to the most prominent $\gamma$ line of $^{133}$Ba ---at 356 keV--- (18.5 h of measurements) and the extracted energy resolution.}
\label{fig:spectra} 
\end{figure}

The energy calibration of the LDs was obtained by exploiting an X-ray peak originated from the X-ray fluorescence of copper, induced by the detector-holder exposition to $\gamma$'s and muons from the environmental radioactivity. The Cu characteristic X-rays feature an intense K$_{\alpha}$ peak with an energy of 8.05 keV, which is visible in the low energy part of the LD measured spectra, as illustrated in figure~\ref{fig:spectra}.

 A test on the LD performance over a broad energy range was performed, as this may provide interesting details on the resolution. For this, a $\gamma$ calibration was carried out by placing a $^{133}$Ba source (with an activity $\sim$1~MBq) outside the cryostat, but at a distance of $\sim$1.5~m from the detectors to minimize pile-up events. The most intense $\gamma$ quanta (7\%--62\% of intensity) emitted by the $^{133}$Ba source have energies of 81, 276, 303, 356, and 384~keV; the 356-keV $\gamma$ line has the highest intensity. Thanks to the comparatively thick Ge wafers used (0.5 mm, e.g. three times thicker than LDs of CUPID-0 \cite{Azzolini:2018tum} and CUPID-Mo  \cite{Armengaud:2020a}), the $^{133}$Ba $\gamma$s are well seen in the calibration data of LDs as illustrated for one of them in figure~\ref{fig:spectra}. In addition to the full-energy peaks, two prominent Compton edges are present in the spectrum. They are at the expected positions (207 and 164 keV) with respect to the two most intense $\gamma$ lines (356 and 303 keV respectively). 
 
 Due to the different positions of the LDs inside the cryostat, we could not see the $\gamma$ lines on LD23 as an internal thick copper cold plate inevitably intercepted the line of sight between this detector and the source. The three other LDs showed an excellent energy resolution on the $\gamma$ lines, in particular the energy resolution of the most intense peak at 356~keV is measured with 0.7--1.8~keV FWHM. The result achieved by the bolometer LD9 and demonstrated in the inset of figure~\ref{fig:spectra}, 0.71(5)~keV FWHM at 356 keV (i.e. 0.1\% RMS), is the best ever obtained in this energy region for any $\gamma$ detector to our knowledge, surpassing the energy resolution of Ge diodes \cite{Belli:2020,Danevich:2020,Danevich:2022,Agostini:2019a}.

\begin{figure}
\centering % \begin{center}/\end{center} takes some additional vertical space
\includegraphics[width=0.7\textwidth]{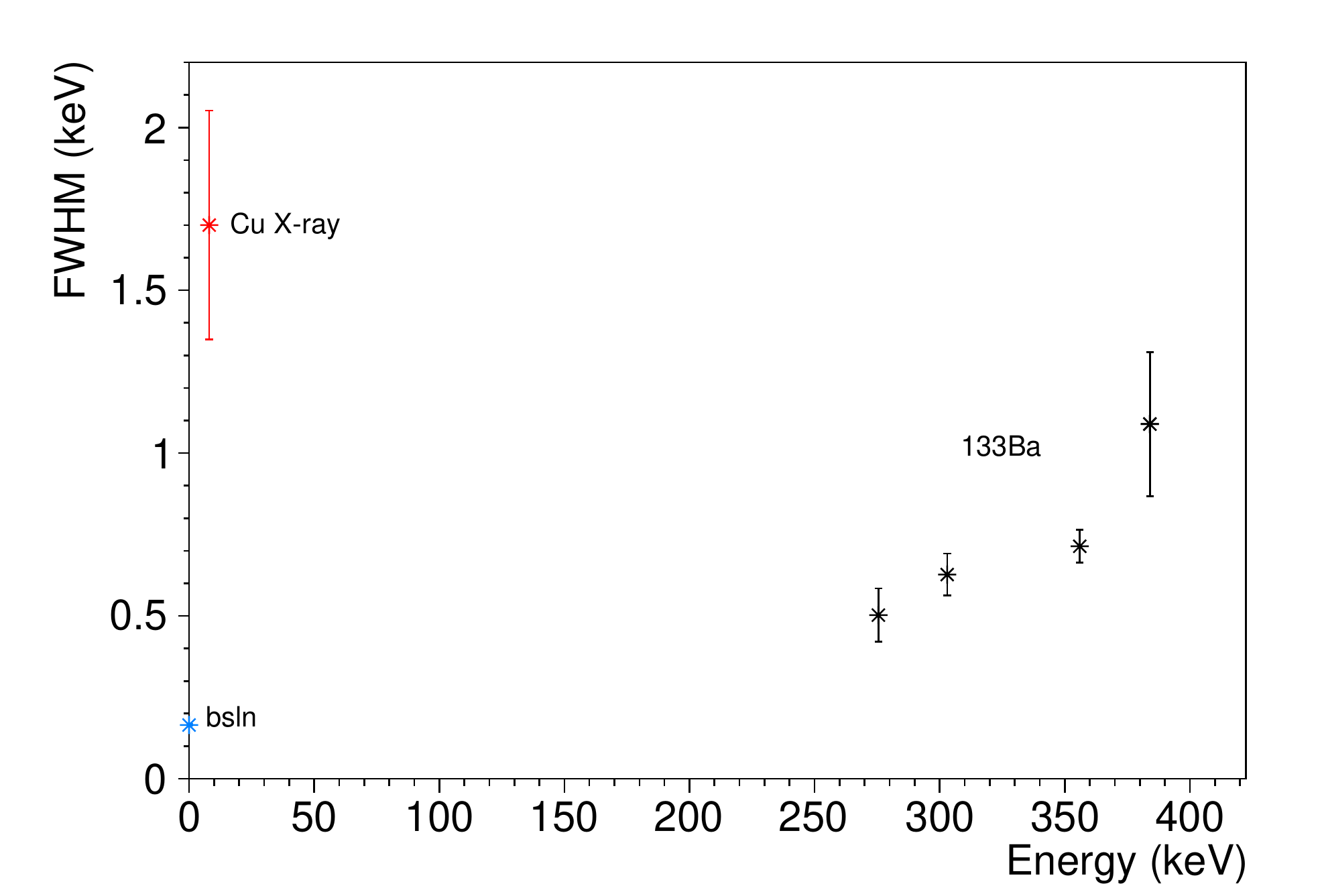}
\caption{LD9 energy resolution (FWHM) measured for baseline noise traces (bsln, in blue), the 8~keV Cu X-rays (in red), and $\gamma$ quanta of $^{133}$Ba detected in the 250--400 keV energy range (in black).}
\label{fig:FWHM}
\end{figure}

The widths of the X-ray peak and the $\gamma$ lines are however much larger than the baseline noise, as illustrated in figure~\ref{fig:FWHM}. Moreover, the resolution of the low-energy X-ray peak (near surface interaction) is about a factor 2 worse than that of more energetic $\gamma$'s (bulk interaction). These effects could be explained by a position dependence of the detector response on the impact point, typical for small and thin absorbers~\cite{Armengaud:2017,Novati:2019,CrossCupidTower:2023a}. Another possible reason of wider peaks is a variation of the detector response in time, related to temperature drifts. This can also explain the observed spread in the energy resolutions ---much higher than the statistical error on the peak width--- as the instabilities can affect the individual devices in different ways. Unfortunately, the heaters were not used in the present experiment, thus, these results can be improved with a more stable cryogenics (in terms of temperature drifts and noise). 

Taking into account the sea level location of the experiment, we can also exploit muons to calibrate LDs, as proposed in \cite{Novati:2019} and taking into account the Ge wafer thickness. A muon-induced distribution of events detected by each LD is illustrated in figure~\ref{fig:spectra}. We remark that the X-ray calibration, the $\gamma$ calibration and the position ---at around 300 keV--- of the bump due to cosmic muons crossing vertically the Ge wafer are coherent when taking into account the detector slight non-linearity. 

Knowing the energy scale of the LDs, we can evaluate the detector sensitivity in terms of a voltage signal (before amplification) per unit of deposited energy, typically given in $\mu$V/keV. The obtained sensitivities of the LDs range between 1.1 and 1.5 $\mu$V/keV, which is good for such NTD-instrumented devices \cite{Beeman:2013b,Artusa:2016,Armengaud:2017,Barucci:2019,Armengaud:2020a}, but not extraordinary due to the necessity of operating NTDs at higher temperatures. 

The baseline widths of the four CUPID LDs, as well as several other operational characteristics of these devices, such as the working points, the time constants of the bolometric signal, the sensitivities and the energy resolutions at 356 keV are summarized in table \ref{tab:LDperformance}.

\begin{table}
\centering
\caption{Performance of four LDs tested in a pulse-tube cryostat. For each light detector we report the working point of the thermistor (NTD resistance $R_{NTD}$ for a given current $I_{NTD}$), characteristic times of rising and descending parts of the bolometric signal (rise- and decay-time parameters, $\tau_{r}$ and $\tau_{d}$), detector sensitivity represented by a signal voltage amplitude per unit of the deposited energy ($A_s$), RMS baseline width (RMS$_{bsln}$) and energy resolution at 356 keV $\gamma$s (FWHM$_{356}$). The last row reports the harmonic mean values of the parameters.}
\smallskip
\begin{tabular}{lccccccc}
\hline
LD$\#$ & $R_{NTD}$ & $I_{NTD}$ & $\tau_{r}$ & $\tau_{d}$ & $A_s$ & RMS$_{bsln}$ & FWHM$_{356}$    \\
~ & (M$\Omega$) & (nA) & (ms) &  (ms) & ($\mu$V/keV) & (eV) &  (keV)  \\
\hline
LD7  & 4.8 & 2.1 & 1.7 & 6.4 & 1.3 & 73(1) & 1.84 $\pm$ 0.16  \\
LD9  & 6.7 & 1.8 & 2.0 & 6.5 & 1.4 & 70(1) & 0.71 $\pm$ 0.05   \\
LD23 & 6.2 & 1.7 & 1.7 & 6.2 & 1.5 & 73(1) & --                \\
LD28 & 3.7 & 2.1 & 1.8 & 9.8 & 1.1 & 89(1) & 0.91 $\pm$ 0.08   \\
\hline
Mean & 5.1 & 1.9 & 1.8 & 7.0 & 1.2 & 76    & \\ 
\hline
\end{tabular}
\label{tab:LDperformance}
\end{table}

The baseline noise width of the filtered data of the LDs was spanning between 70 eV to 90 eV RMS, which is typical for such devices equipped with an NTD sensor, used in several previous experiments~\cite{Armengaud:2020a,Armengaud:2017b,Artusa:2016,Armatol:2021a,Alfonso:2022,Poda:2021}.
Unlike most of these searches, CUPID has an open detector structure \cite{Alfonso:2022} and does not use an optical reflector around the absorber. This is indispensable for surface background rejection, as already discussed. This decreases light collection that results in about twice lower signal in the LD compared to detectors surrounded by a reflective film \cite{Armengaud:2020a,Armatol:2021a,CrossCupidTower:2023a}. This requires high performance LDs for $\alpha$ rejection, that translates in the CUPID goal of 100 eV RMS for the baseline width. This is needed to reject about 99.9\% of $\alpha$ particles preserving about 90\% of $\gamma$($\beta$) events \cite{Armstrong:2019inu}. The four LDs tested here meet this requirement, as stated above and shown in table~\ref{tab:LDperformance}. It is important to stress once more that this test was performed in a pulse-tube cryostat, where vibrational noise has an impact on the LD performance.

%%============================================================================================
\section{Conclusions}

In this paper a description of the CUPID baseline LD assembly was covered, in addition to the first results on the performance of four LDs tested in a pulse-tube cryostat. The RMS baseline resolution showed to comply with the CUPID goal (100 eV RMS) with good reproducibility (70, 73, 73 and 89 eV RMS). Detector rise-times ($\sim$2 ms) must be shortened to achieve the CUPID objective, but this can be obtained by reducing the operation resistance of the thermistors ---acting on the geometry and/or on the doping level--- and by improving the thermal connection between the wafer and the NTD. Excellent energy resolutions were observed in a $\gamma$ calibration, that was performed to validate the energy scale of the detectors. In particular, one LD exhibited an energy resolution of 0.71(5) keV FWHM at 356 keV, which is the highest ever achieved for any $\gamma$ detector at that energy.  Currently, a validation of a full CUPID tower containing 30 LDs (including the four here characterized) and 28 Li$_2$MoO$_4$ crystals, is ongoing in the Gran Sasso underground laboratory in Italy.

%%============================================================================================
\acknowledgments

The cryostat used for the tests here described ---installed at IJCLab (Orsay, France)--- was donated by the Dipartimento di Scienza e Alta Tecnologia of the Insubria University (Como, Italy). This work was supported by the Agence Nationale de la Recherche, France (CUPID-1, Projet-ANR-21-CE31-0014); by the Institut National de Physique Nucl\'eaire et de Physique des Particules (IN2P3); by the Istituto Nazionale di Fisica Nucleare (INFN); by the European Research Council (ERC) under the Marie Sklodowska-Curie Grant Agreement no. 754496; by the Italian Ministry of University and Research (MIUR) through the grant Progetti di ricerca di Rilevante Interesse Nazionale (PRIN 2017, Grant no. 2017 FJZMCJ); by the US National Science Foundation under Grant nos. NSF-PHY-1401832, NSF-PHY-1614611, and NSF-PHY-1913374. This material is also based upon work supported by the US Department of Energy (DOE) Office of Science under Contract nos. DE-AC02-05CH11231 and DE-AC02-06CH11357; and by the DOE Office of Science, Office of Nuclear Physics under Contract nos. DE-FG02-08ER41551, DE-SC0011091, DE-SC0012654, DE-SC0019316, DE-SC0019368, and DE-SC0020423. This work was also supported by the National Research Foundation of Ukraine under Grant no. 2020.02/0011 and by the National Academy of Sciences of Ukraine in the framework of the project ``Development of bolometric experiments for the search for double beta decay'', the grant number 0121U111684. This research used resources of the National Energy Research Scientific Computing Center (NERSC). The CUPID collaboration acknowledges the scientific and technical contributions from all collaborators. More information on the technical details of the experiments and the collaboration policies can be found at https://cupid.lngs.infn.it.

%%============================================================================================

%\bibliographystyle{JHEP}
%\bibliography{Bibliography} 

\end{document}